\documentclass[floatfix,twocolumn,showpacs,preprintnumbers,amsmath,amssymb]{revtex4}
\usepackage{amsmath,amssymb,color,epsfig,latexsym,natbib,graphicx,dcolumn}
\newcommand{\bm}{\mathbf}
\bibpunct{[}{]}{,}{n}{,}{,}             
\def\DEL#1{{\textcolor{green}{}}}         

 \newcommand{\bB}{\bf{B}}

 \newcommand{\rem}[1]{}

\newcommand\vecp[1]{\vec{#1}}                   

\newcommand{\be}{\begin{equation}} \newcommand{\ee}{\end{equation}}

\def\bB0{\vecp{B}_0}

\begin{document}
\title{\bf A paradigmatic flow for small-scale magnetohydrodynamics:  properties of the ideal case and the collision of current sheets}
\author{E. Lee$^{1,2}$, M.E. Brachet$^{1,3}$, A. Pouquet$^1$, P.D. Mininni$^{1,4}$, and D. Rosenberg$^1$}
\affiliation{(1) Geophysical Turbulence Program, National Center for Atmospheric Research, P.O. Box 3000, Boulder, Colorado 80307-3000, U.S.A.}
\affiliation{(2) Department of Applied Physics and Applied Mathematics,
Columbia University, 500 W. 120th Street, New York NY 10027, U.S.A.}
\affiliation{(3)  \'Ecole Normale Sup\'erieure, 24 Rue Lhomond, 75231 Paris, France}
\affiliation{(4) Departamento de F\'\i sica, Facultad de Ciencias Exactas y
         Naturales, Universidad de Buenos Aires, Ciudad Universitaria, 1428
         Buenos Aires, Argentina}

\begin{abstract}
We propose two sets of initial conditions for magnetohydrodynamics (MHD) in which both the velocity and the magnetic fields have spatial symmetries that are preserved by the dynamical equations as the system evolves. When implemented numerically they allow for substantial savings in CPU time and memory storage requirements for a given resolved scale separation.
 Basic properties of these Taylor--Green flows generalized to MHD are given, and the ideal non-dissipative case is studied up to the equivalent of $2048^3$ grid points for one of these flows. The temporal evolution of the logarithmic decrements $\delta$ of the energy spectrum remains exponential at the highest spatial resolution considered, for which an acceleration is observed briefly before the grid resolution is reached. Up to the end of the exponential decay of $\delta$, the behavior is consistent with a regular flow with no appearance of a singularity. The subsequent short acceleration in the formation of small magnetic scales can be associated with
a near collision of two current sheets driven together by magnetic pressure. It leads to strong gradients with a fast
rotation of the direction of the magnetic field, a feature also observed in the solar wind.
\end{abstract}
\maketitle
\section{Introduction}
Astrophysical and geophysical flows are highly turbulent, with strong coupling between a wide range of spatial and temporal scales. The complex behavior of such flows is far from understood, and their study through direct numerical simulation (DNS) in three space dimensions is limited to modest scale separation, even at the largest resolution achieved today \cite{kaneda03,1536ab}.  In fact, many numerical experiments designed to study turbulent flows are confined to Cartesian geometry and periodic boundary conditions in order to maximize the scale separation (or the Reynolds number in the dissipative case) in the computations.  Furthermore, scaling laws of turbulence parameters, such as the inertial index of the energy spectrum, the Kolmogorov constant, or the skewness and flatness of velocity derivatives, all converge slowly to their high Reynolds number limits (assuming they exist), and they converge at different speeds \cite{kaneda06}.

There are several ways to obtain a larger scale separation in simulations of hydrodynamics and magnetohydrodynamics (MHD).  Adaptive mesh refinement (AMR) is a method of improving resolution locally, based on structures that develop in the flow, without having to refine the grid everywhere.  For dissipative turbulence, however, AMR can be difficult or costly to implement, unless the emerging structures are strongly concentrated in space \cite{duane_amr}.  Success with AMR has been achieved in certain contexts, such as in the study of the possible development of singularities in hydrodynamics and MHD \cite{grauer1,grauer2}.

In more homogeneous flows, an alternative is to model the small scales that are not explicitly resolved in a simulation.  Such methods abound for neutral fluids (see e.g., \cite{lesieur}), but in the case of MHD the research is less developed \cite{julien}.  An additional difficulty in MHD is the lack of laboratory experiments at large magnetic Reynolds number to evaluate such models against DNS and thus to further improve them \cite{holm02}.  Also, although they are useful for many applications, these methods cannot be used to study mathematical problems such as the development of singularities in finite time, since the original equations are, as a rule, modified or replaced by a different set of equations to mimic the effect of the turbulent mixing and dissipation at unresolved scales.

A third approach is to construct a flow with symmetries that are hypothesized to hold under the evolution of the relevant dynamical equations.  Numerical simulations of such flows have been performed using codes designed to exploit their symmetries, leading to significant advances in the study of turbulent flows \cite{brachet90,brachet91}, due to substantial savings in both computing time and memory (see also \cite{kida}).  Like AMR, flows with symmetries are suitable for studying the possible development of singularities in an ideal flow \cite{cichowlas_brachet}.  Efficient codes implementing flow symmetries have also been utilized in the study of MHD flows, although in the context of a forced turbulent dynamo rather than singularity \cite{kida_dynamo,nore97}.  It is worth noting that in both of these cases, although the magnetic field was evolved by a code employing symmetries, it was initialized as a small, random seed field whose symmetries were incidental and not deterministic; therefore, only a symmetric velocity field needed be prescribed.

In this paper we propose two paradigmatic MHD flows with symmetries.  Numerical implementation of these symmetries are used to demonstrate that the symmetries are preserved by the MHD equations as the system evolves for many turnover times, as evinced by a comparison of simulations with and without imposed symmetries.  As the implemented symmetries allow for substantial savings in memory and computing time, we are able to study the evolution of an ideal MHD flow at unprecedented spatial resolution.  The structure of the paper is as follows.  In the next section we present the equations and definitions of the possible flows that can be studied in this context, giving as well some of their basic geometrical properties. Section \ref{s:result} is devoted to a study of the ideal non-dissipative case and to the rapidity of the development of small scales through an analysis of the temporal evolution of the analyticity strip (to be defined below).  At the end of Section \ref{s:result} we also discuss numerical simulations showing that the symmetries of the flow are preserved for all times considered in the present work.  Finally, Section \ref{s:conclu} presents our conclusions.

\section{Taylor--Green flows} \label{s:TG}

\subsection{The MHD equations}

We recall here the ideal MHD equations for an incompressible flow; these equations stem from the conservation of momentum and from Maxwell's equations for velocities substantially less than the speed of light. In terms of a dimensionless velocity $\bm v$ and magnetic induction $\bm b$, they are
\begin{eqnarray}
\partial_t {\bm v} + \bm v \cdot \nabla \bm v &=& -\nabla P + \bm j \times \bm b  \ , \\ 
\partial_t {\bm b} &=& \nabla \times (\bm v \times \bm b)  \ ,  
\end{eqnarray}
together with $\nabla \cdot \bm v = 0 = \nabla \cdot \bm b$; $\bm j=\nabla \times \bm b$ is the current density and $P$ the pressure.
Note that ${\bf b}$ is in fact expressed in terms of an Alfv\'en velocity. 
The total energy 
\begin{equation}
E=\frac{1}{2} \langle {\textbf{v}}^2+\bm b^2 \rangle =E_V+E_M ,
\end{equation}
(where $E_V$ is the kinetic energy, $E_M$ is the magnetic energy, and $\langle \ \rangle$ signifies integration over the domain), the cross-correlation
\begin{equation}
H_C=\frac{1}{2} \langle \bm v \cdot \bm b \rangle ,
\end{equation}
and the magnetic helicity 
\begin{equation}
H_M= \frac{1}{2} \langle \textbf{a} \cdot \textbf{b} \rangle ,
\end{equation}
(where $\bm b = \nabla \times \bm a$, and $\bm a$ is the vector potential) are conserved by these equations in three space dimensions. The conservation of these quantities (up to the numerical truncation error) is monitored in the simulations and used as one criterion to decide the last reliable time for which the accuracy of the solutions is satisfactory.

\subsection{Hydrodynamic Taylor--Green flows}

The Taylor--Green (hereafter, TG) vortex is an important hydrodynamic flow for the insights it provides at moderate complexity but low cost. 
It mimics von K\'arm\'an flows between two counter-rotating disks, as used in several turbulence experiments (including recent experiments to reproduce generation of magnetic fields by dynamo action; see e.g.\cite{bourgoin02} and references therein). It was originally motivated as an initial condition that, in spite of its symmetries, would lead to the rapid development of small spatial scales \cite{Taylor37} and therefore proposed as a paradigm of the direct cascade of energy in a turbulent flow. Analytic expansions were considered to study the possible development of singularities in this flow \cite{Morf80}.  Fundamental to the TG vortex in later numerical studies were its symmetry properties that allowed for gains of a factor of four in linear resolution \cite{brachet83,Pelz85,mebk2,cichowlas_brachet}, a feature that was also exploited to study the generation of magnetic fields by dynamo action \cite{nore97}.

The simplest TG vortex is given by the velocity field initial data 
\begin{eqnarray}
v_x &=& v_0 \sin x \cos y \cos z \\
v_y &=& -v_0 \cos x \sin y \cos z \\
v_z &=& 0 .
\label{eqn:simple_vTG}\end{eqnarray}
The $z$-component of the velocity grows with time as the system evolves, due to the pressure gradient in the momentum equation.  When the hydrodynamic equations are integrated in time starting from this initial condition, the symmetries of the flow are preserved for many turnover times.

Within a periodic cube of length $2\pi$, there are two planes of mirror symmetries (or anti-symmetries)
in each dimension: $x=0,\pi$; $y=0,\pi$; and $z=0,\pi$. The flow is also invariant by rotation of $\pi$ around the two axis $x=z=\pi/2$ and $y=z=\pi/2$ and by rotation of $\pi/2$ around the third axis $x=y=\pi/2$. As a consequence of the last rotational invariance, note that $v_y(x,y,z)=-v_x(\pi-y,x,z)$.
Exploiting these symmetries allows one to recover the dynamics of the entire domain from computations in a ``symmetry'' box of size $[0,\pi/2]^3$ \cite{brachet83,nore97}. By reducing the amount of information needed to represent the full cube, the symmetries allow for simulations at much higher Reynolds numbers, or simulations of the ideal inviscid flow for a longer time than a full DNS for a given fixed cost.   Specifically, at a given computational cost, the ratio of the largest to the smallest scale available to a computation with enforced Taylor--Green symmetries is enhanced by a factor of 4.  Equivalently, at a given Reynolds number, a factor of 4 in savings in linear resolution offered by the Taylor--Green flow thus leads to a factor of 32 savings in total computational time and a factor of 64 savings in memory usage.

\subsection{Two possibilities in MHD}  
With the velocity given above, there are several possibilities for the choice of initial conditions for the
magnetic field with the same symmetries. The first set of magnetic initial conditions will be called the ``insulating'' case, with $\bm b$ everywhere normal to the walls, and the current $\bm j$ contained within what can be called the ``insulating box'' $[0,\pi]^3$; hence, the walls in this box are insulating, with the magnetic field behaving like the vorticity in the hydrodynamic TG case:
\begin{eqnarray}
b_x^{(i)} &=& b_0 \cos x \sin y \sin z  \\
b_y^{(i)} &=& b_0 \sin x \cos y \sin z  \\
b_z^{(i)} &=&-2 b_0 \sin x \sin y \cos z  .
\label{eqn:btg_I}\end{eqnarray}

The second possibility in MHD is called the ``conducting'' case; it has the magnetic field, just like $\bm v$,  contained entirely within the box:
\begin{eqnarray}
b_x^{(c)} = b_0 \sin2x\cos2y\cos2z \ , \\
b_y^{(c)} = b_0 \cos2x\sin2y\cos2z \ , \\
b_z^{(c)} = -2 b_0 \cos2x\cos2y\sin2z \ .
\label{eqn:btg_C}\end{eqnarray}

For this flow $H_C\equiv 0$ (uncorrelated $\bm v$ and $\bm b$), and it yields a current that is everywhere perpendicular to the walls of the conducting $[0,\pi]^3$ box.  In this paper, we concentrate on the study of the first flow, which we name IMTG (``insulating magnetic Taylor--Green'') hereafter.

\subsection{Basic properties of the IMTG flow}
The initial velocity in the insulating box has a shear layer between two counter--rotating eddies, and
the magnetic field is initially proportional to the vorticity.  Local kinetic and magnetic helicities are {\it anti-symmetric} with respect to the planes of mirror symmetries and thus vanish globally.  Kinetic helicity, which is defined as $H_K=\left< {\bm v} \cdot \mbox{\boldmath $ \omega $} \right>/2$, is not generally conserved by the ideal MHD equations.

Noting that $E_V=v_0^2/8$, $E_M=3 b_0^2/8$, $H_C=0$ and $H_M=0$, in what follows we take $v_0=1$ and $b_0=1/\sqrt{3}$ (and thus $E_V=E_M=1/8$ at $t=0$).  Fourier space components of the symmetric fields are non-zero only for jointly even or jointly odd wavenumbers.  Even and odd components are treated separately in the code, using specially designed Fourier transforms that are performed on $N/4$ points, where the resolution $N$ denotes the number of grid points in the full periodicity box.  Computations are dealiased using the $2/3$ rule, and the wavenumbers are thus integers in the range [$k_{min}=1, k_{max}=N/3$].  Finally, time differencing is achieved using, in general, a second-order Runge-Kutta method.

When the IMTG flow is used as the initial condition for a freely decaying simulation, the symmetries are preserved by the time evolution for many turnover times. This is shown in the next section for the ideal case (of interest for the present work), and will be shown for the viscous case in a forthcoming work.

\section{Numerical results} \label{s:result}
\subsection{Temporal evolution in the ideal case}
In the ideal case, $E$, $H_M$, and $H_C$ are conserved (except for time-stepping errors), and the flow, with a finite number of modes, eventually evolves to statistical equilibria that depend on these conserved quantities \cite{stat_mech}.  In the simplest case, equipartition between modes obtains, leading to an accumulation of energy in the smallest scales with a $k^{d-1}$ spectrum in $d$ space dimensions, the so-called ultra-violet energy catastrophe.  It was shown recently \cite{mebk2} that for $d=3$, the small-scale $k^2$ spectrum acts as an eddy viscosity to the modes at large scales, thereby putting on a strong footing the assertion that the ideal problem contains the dynamics of the dissipative problem, at least for times short enough that the equipartition has not yet been able to reach all scales.
 
 \subsection{The singularity test}
 In this light, it is of interest to examine the behavior of the IMTG flow we just proposed. One central question is whether ideal MHD flows develop singularities in a finite time. In a celebrated paper \cite{BKM} 
 it was shown that, for the Euler equation,  the $L^\infty$ norm of vorticity controls the breakdown of solutions; this result was extended to MHD, in which case the singularity is controlled by the combined $L^\infty$ norm of vorticity and current \cite{klapper96, CKS}. 
 There are a number of carefully constructed numerical studies devoted to ideal behavior for both Euler and for ideal MHD, with no clear-cut conclusion of a specific flow showing signs of singularity
 (see \cite{kerr,grauer1,hou} for the Euler case and 
 \cite{ideal_2d,grauer2,kerr_brandenburg} for MHD); note that the
particular choice of initial conditions is essential, as exemplified in \cite{kerr}. 

\begin{figure} 
\includegraphics[width=8.6cm]{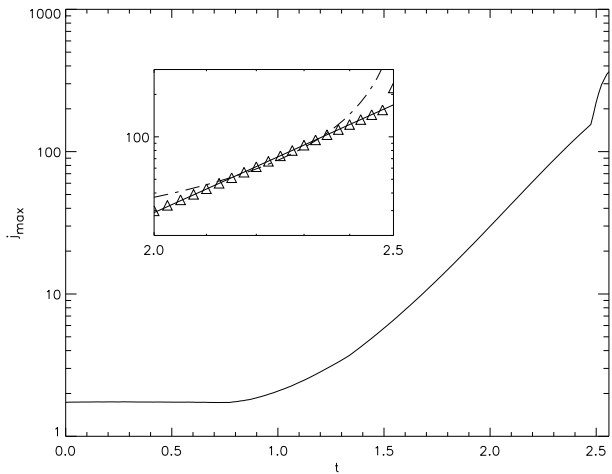}\\
\caption{Maximum of the current in semi-log in a $2048^3$ IMTG run; the insert gives a log-log zoom, with triangles for data, dash-dot for the fit with $[t_*-t]^{-1}$, and solid line for a power law. }
\label{fig_jmax} \end{figure}

\begin{figure} 
\includegraphics[width=8.6cm]{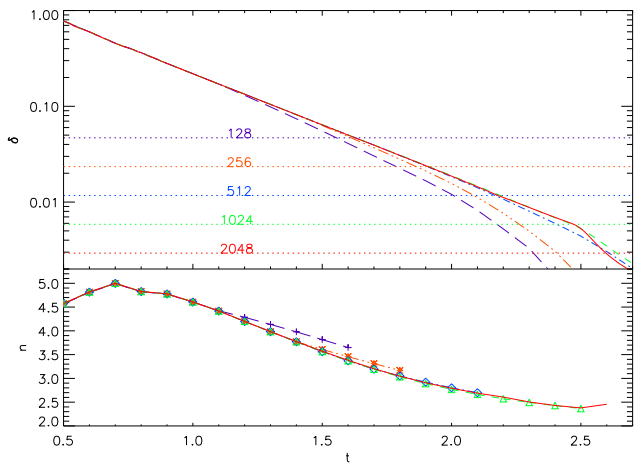}\\
\caption{(Color online) {\it Top:} Logarithmic decrement of the total energy spectrum for various resolutions in the $128^3$ (long dash), $256^3$ (dash-triple dot), $512^3$ (dash-dot), $1024^3$ (dash) and $2048^3$ (solid) runs. Note the acceleration in $\delta$ in the $2048^3$ run at $t\approx 2.5$ {\it Bottom:} Spectral index in the $128^3$ ($+$), $256^3$ ($*$), $512^3$ ($\diamond$), $1024^3$ ($\bigtriangleup$) and $2048^3$ (solid line) runs.}
\label{fig_delta_n} \end{figure}

In this context, we show in Fig.\ \ref{fig_jmax} the evolution of the maximum of the current $j_{max}$ in semi-log coordinates in a run computed on an equivalent grid of $2048^3$ points. 
An approximately exponential phase up to $t \approx 2.47$ is followed by a more abrupt growth corresponding to a change in the spatial location of $j_{max}$.  The exponential behavior at early times has been observed by a number of authors, and can be related to a shear-induced instability.  In fact, an exponential fit is particularly good in the range $t=1.80$ to $t=2.27$, where the least square error $\xi \approx 8 \times 10^{-3}$. A power-law fit also appears appropriate in the range $t=2.10$ to $t=2.47$, with now a least-square error  $\xi \approx 2 \times 10^{-2}$. The inset in Fig.\ \ref{fig_jmax} shows a blow-up of the evolution of $j_{max}$, in log-log coordinates, with this power-law fit up to the sharp transition at $t\approx 2.5$. In spite of the steepness of the index for the power-law fit ($j_{max} \sim t^{7.9}$), the error for the power-law fit is about four orders of magnitude smaller than for a $j_{max} \sim [t_*-t]^{-\alpha}$ fit in the same range, as proposed in \cite{BKM,klapper96,CKS} for a singularity to occur at a finite time $t_*$. The inset in Fig.\ \ref{fig_jmax} also shows the best fit to this expression with $\alpha=1$.

The resolution of the run does not allow us to compute far beyond the time shown. Indeed, from a visual inspection of the flow, the computation can be considered reliable up to $t \approx 2.61$, at which time aliasing ripples in the current appear, roughly at the 15\% level (not shown). As a comparison, the level of aliasing in the current at $t \approx 2.55$ is 9\% (below we discuss two different accuracy criteria based on the evolution of the energy spectrum and the conservation of the MHD invariants). We are thus led to conclude that, due to the 
substantially smaller errors of an exponential or power-law fit than for a $[t_*-t]^{-\alpha}$ fit to the $L^\infty$ norm of the current or vorticity, the flow shows no sign of developing singularities for the reliable duration of the computation. Additional studies will be needed to examine the later temporal evolution of the ideal flow at higher resolution.

\subsection{Logarithmic decrement and accuracy criteria}

Another way to determine whether a flow remains regular at all times or develops a singularity is to
measure the logarithmic decrement. This quantity is also useful as an accuracy criterion for gauging up to what time the ideal computation is reliable. The logarithmic decrement is defined through a (logarithmic) fit of the total energy spectrum $E(k,t)$ [defined in the usual way, with $\int E(k,t)dk=E(t)$] with the expression \cite{sulem}:
\begin{equation}
E(k,t) =  C(t) k^{-n(t)} e^{-2\delta(t)k} \  .
\label{decrement2}\end{equation}
The logarithmic decrement $\delta$ must 
not reach zero (e.g. by decaying exponentially in time, $\delta(t) = \delta_0 e^{-t/\tau}$) for the flow to remain regular at all times. Independently of its time dependence, the instantaneous value of $\delta$ can also be compared with $\Delta x$ (the distance between spatial grid points) to quantify the accuracy of the computation.

We show in Fig.\ \ref{fig_delta_n} the temporal evolution of $\delta$ computed on the total energy spectrum and for simulations with resolutions ranging from $128^3$ to $2048^3$ grid points. A striking result that appears only in the highest-resolved run, again at $t \approx 2.5$, is an acceleration in the decrease of $\delta$.  This sharp change in the evolution of $\delta$ is observed when the fit given by Eq.\ (\ref{decrement2}) is done using the total energy spectrum or the magnetic energy spectrum from the simulation, but not when using its kinetic counterpart.  These results were verified to be independent of the timestep used (which was halved twice), as well as of the order of the temporal scheme (which was tested at second and fourth order).
 
Interestingly, the accelerated decrease of $\delta$ coincides with the accelerated increase of $j_{max}$ already observed.  The change in $\delta$ can be associated with a concurrent change in the functional form of the magnetic energy spectrum. After $t \approx 2.5$, the energy spectrum in the $2048^3$ run shows a slight excess of energy at small scales compared with Eq.\ (\ref{decrement2}).  Indeed, the functional form for the energy spectrum given in Eq.\ (\ref{decrement2}) may become too simple after $t\approx 2.5$.  We have also verified that the acceleration in the decrease of $\delta$ persists when the range of wavenumbers used for the fit is changed.

To evaluate the credibility of the computation, we set as our criterion for accuracy $\delta k_{max}\ge 2$, where $\delta(t)$ is fit over the range $k \in [2,400]$ for the ($2048^3$) run, which has maximum wavenumber $k_{max}=N/3=682$ after dealiasing.  By this measure, the computation is reliable up to $t\approx 2.61$, in agreement with the above examination of aliasing levels in the current density in configuration space.  For each run, the threshold of reliability based on the $\delta k_{max}\ge 2$ criterion is marked by a dotted horizontal line in Fig.\ \ref{fig_delta_n}; that is, each line (labeled by its spatial resolution) corresponds to the minimum value of $\delta$ that can be resolved by each simulation.  Therefore, the point at which $\delta(t)$ intersects the horizontal cutoff line for the given resolution marks the last reliable time.  It is critical to note that the $2048^3$ run shows the acceleration in the evolution of $\delta$ taking place well before the cutoff time, while the other runs show no indication of acceleration prior to their respective cutoff times.

A third criterion for ensuring accuracy of the computations is the conservation of the invariants of the MHD equations. The time at which conservation breaks down in the simulations is in good agreement with the time obtained from the two other criteria described above.  To give an example, we note that in the $2048^3$ run the energy conservation begins to break down (at the level of $5 \times 10^{-5}$ of the initial energy) at $t\approx2.6$, which is consistent with the time at which $\delta$ reaches the smallest resolved scale of the flow.

Finally, the evolution of the spectral index $n$ in Eq.\ (\ref{decrement2}) as a function of time is shown in Fig.\ \ref{fig_delta_n}, for each run up to its respective cutoff time.  For all resolutions we observe that the spectral index is smaller than in the Euler case \cite{cichowlas_brachet}, leading to a shallower spectrum and likely corresponding to sharper structures in the MHD flow.  This could also be related to the presence of a magnetic field that provides a slower time scale for the spectral energy transfer \cite{1536ab,yoshida07,MHDspec}.

\begin{figure}
\includegraphics[width=8.3cm]{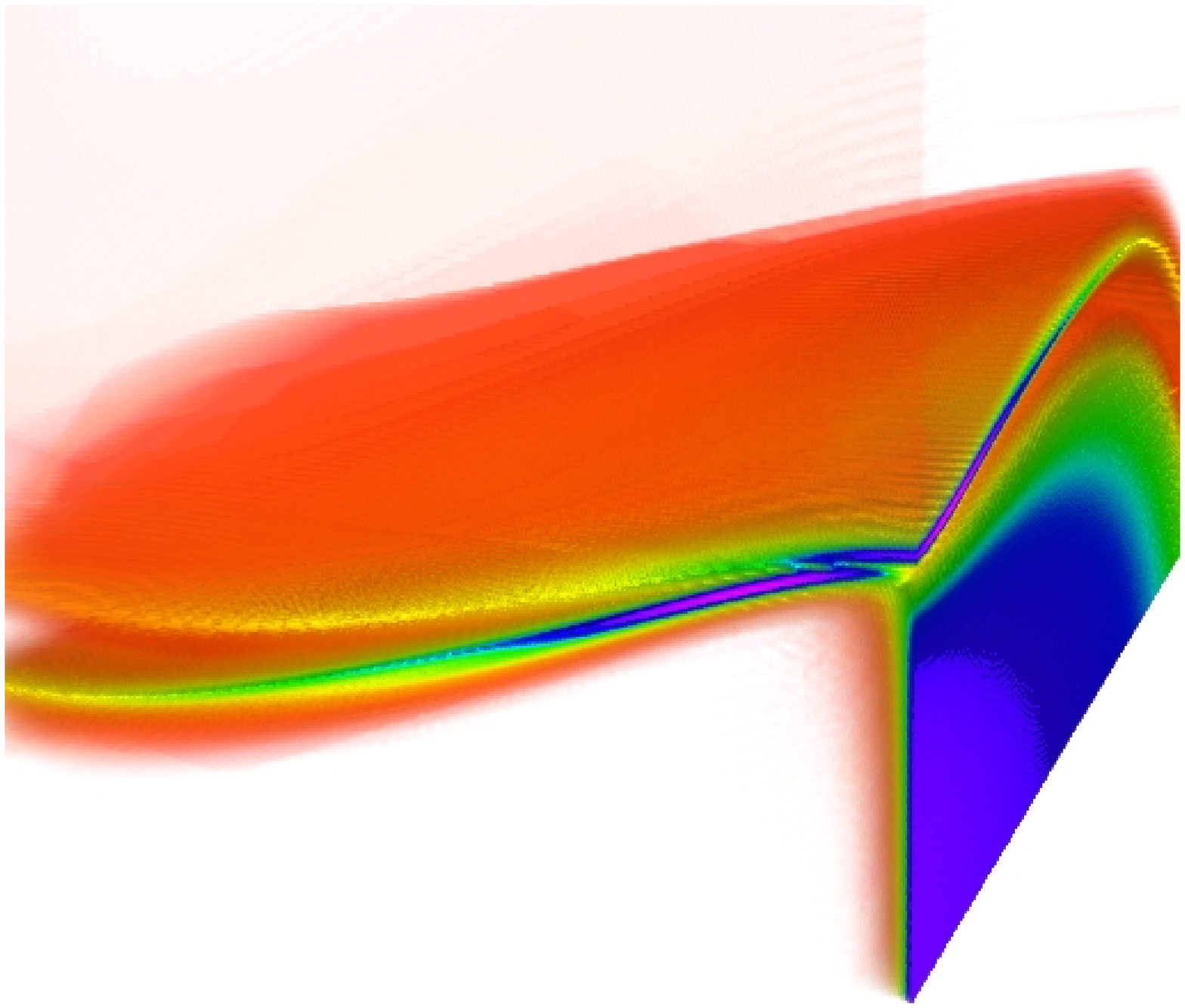}\\
\vskip0.01truein
\includegraphics[width=8.3cm]{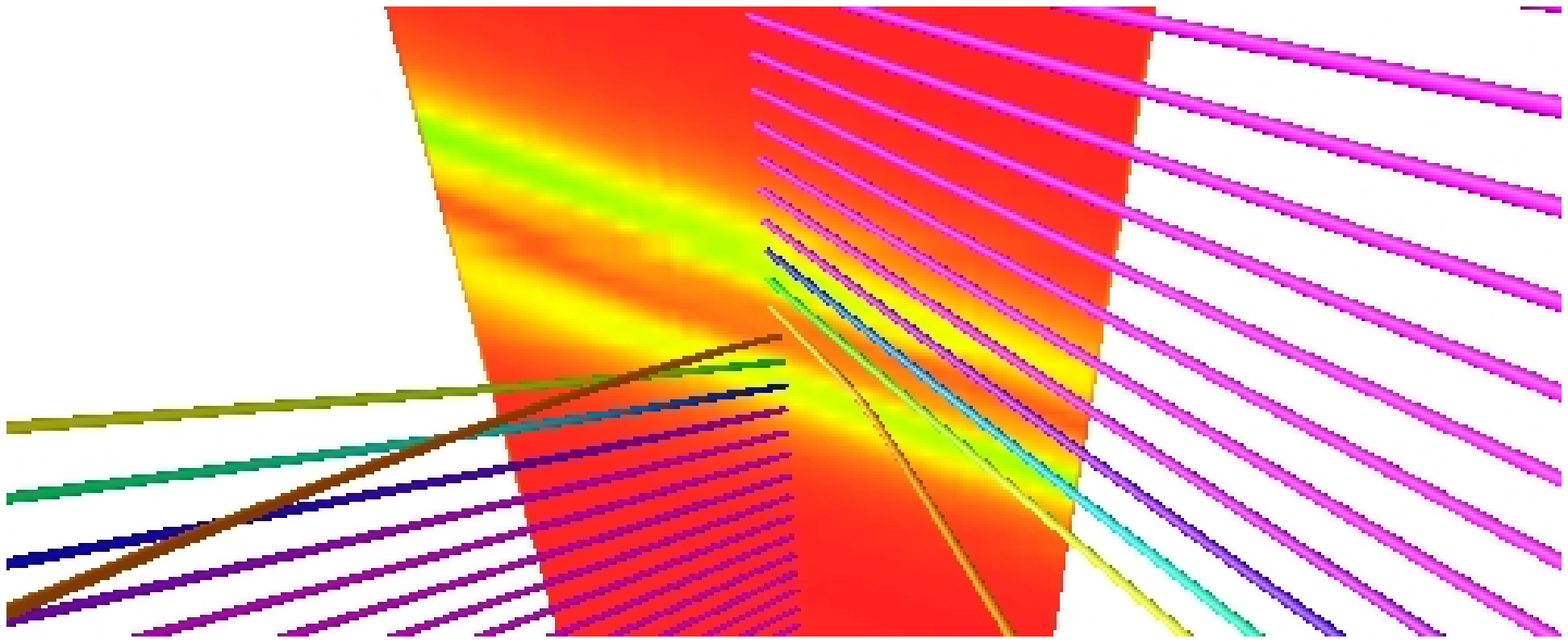}
\caption{(Color online) {\it Top:} Current density intensity in a $200\times 200\times 120$ subvolume of the $2048^3$ run at $t=2.5$. Note the two current sheets approaching each other. {\it Bottom:} Magnetic field lines in the same structure (viewed from the back). The current intensity in a slice is given as a reference.  To the right and left of the slice, the magnetic field is strong (purple color), whereas in the transition region it decreases to $\approx 1/6$ of the maximum (yellow), corresponding to a strong local drop in intensity. Note also the rapid spatial rotation of the field lines between the two current sheets.}
\label{f3} \end{figure}

\subsection{Spatial structures in the ideal case} 

We now turn to an examination of the structures that develop, concentrating on the magnetic field and current in the $2048^3$ run, for which the temporal evolution shows an abrupt change at $t\approx 2.5$.
Because of the symmetries, the strongest gradients tend to develop near the insulating walls, where thin current sheets appear. The roll-up of these sheets, as observed in a dissipative simulation at high resolution of a different flow  \cite{1536ab} does not occur here; presumably the reason is that the conditions for a Kelvin--Helmholtz-like instability are not present, due to the symmetries satisfied by the vorticity and the current.

Two strong current sheets appear near the edges of the box, approximately halfway up along one direction ($z \approx \pi/2)$, whose geometry and evolution we describe here.  Fig.\ \ref{f3} shows the structures in a subvolume with approximately $200\times 200\times 120$ grid points, as well as the magnetic field lines, colored according to their strength.  The current intensity is concentrated in the two thin sheets, which are connected to two other thin sheets running along the insulating walls (only one of which is shown in the figure).  At each of the edges of the box, the two horizontal sheets are closest near the edge and farther apart towards the interior.  The magnetic field around these structures is smooth and large-scale, strong outside the sheets but weak between the two sheets; the magnetic pressure gradient thus provides a force that moves the sheets closer together.  Each individual sheet corresponds to a sheared magnetic field configuration, and the field between the sheets undergoes a rapid rotational shear.

As time evolves, two distinct processes take place: each sheet becomes thinner, and the two sheets continue to be pushed together by the magnetic pressure; these processes occur at different rates. Before $t\approx 2.5$, the thickness of the sheets $e_{\delta}$ decreases faster than the separation distance $e_\Delta$ between the two. The magnetic field lines between the sheets rotate smoothly from the direction observed above the top current sheet to the direction below the bottom current sheet.  At $t\approx 2.5$, $e_{\delta}$ and $e_\Delta$ become comparable, and subsequently the rate of near-collision of the two sheets surpasses the rate at which each sheet thins, hence the acceleration in the evolution of the logarithmic decrement $\delta$ and in the growth of $j_{max}$.  As stated previously, the computation has to be stopped when either $e_\delta$  or $e_\Delta$ becomes comparable to the distance between grid points $\Delta x$.

The rotational shear intensifies as the sheets approach each other.  Indeed, each magnetic field line in Fig.\ \ref{f3} is separated by two grid lengths in the cross section.  At the time shown in the figure ($t=2.5$), a sharp spatial rotation of the field lines of approximately $\pi/2$ can be seen to occur over a short distance, corresponding to a few field lines (about 8 grid lengths), while the drop in the amplitude of the magnetic field between the two sheets occurs over about twice that distance.  The transition region becomes thinner rapidly after $t=2.5$, pushed by the larger magnetic pressure outside the two sheets, and by the time the simulation is stopped the magnetic field rotates by $\pi/2$ almost abruptly.

In comparison, structures in the vorticity are smoother. This is consistent with the fact that the acceleration of $\delta$ is not observed when the logarithmic decrement is computed from the kinetic energy spectrum.

\subsection{Comparison with a run without imposed symmetries}

One question to be addressed is whether the imposed symmetries alter in substantial ways the dynamics of the flow. Or, in other words, if the symmetries of the initial conditions are preserved by the time evolution when symmetries are not imposed in the simulation (at least for times long enough to justify the savings in memory and computing time in free decaying simulations). It is known for example that imposing symmetries can slow the growth of the MHD dynamo instability for some flows \cite{nore97}, in which the fastest growing mode does not obey the imposed symmetries.  To quantify the effect of imposing the symmetries in the evolution of the IMTG flow, we performed two simulations starting from the same initial conditions and with a resolution of $512^3$ grid points: one with imposed symmetries, and another without any imposed symmetry.

\begin{figure} 
\includegraphics[width=8.6cm]{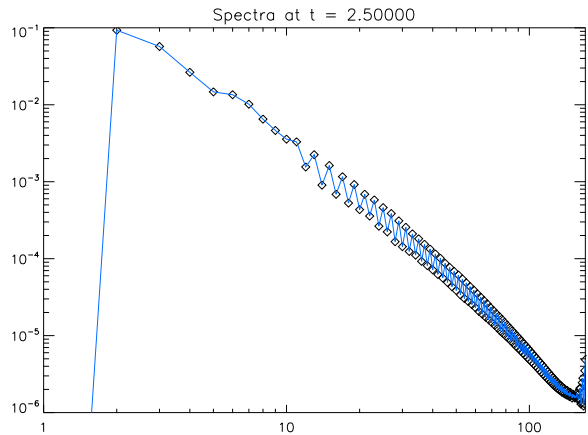}\\
\caption{(Color online) Total energy spectrum for the IMTG flow on a grid of $512^3$ points in the ideal case; results are obtained using a symmetric code  (diamonds) and a run with no imposed symmetries (solid line). No difference is observed except, at a low energetic level, in the lowest mode of the computation. }
\label{fig_512} \end{figure}

Fig.\ \ref{fig_512} shows the energy spectrum near the end of the computations ($t = 2.5$).  Diamonds correspond to data produced with the symmetric code, and the solid line corresponds to the equivalent run with a general code, which allows any non-symmetric mode to evolve freely.  No difference is observed except at the lowest mode and at an energetic level close to round-off errors.  The amplitude of any of these modes can be used to estimate how much energy is contained in modes that do not satisfy the symmetries. In particular, the energy in the modes in the $k=1$ shell: ${\bm k} = (1,0,0)$, $(0,1,0)$, and $(0,0,1)$, which do not satisfy the symmetries of the IMTG flow, in the run without imposed symmetries is $3 \times 10^{-16}$, of the order of the truncation error.  Note that oscillations between even and odd wave numbers are visible in the energy spectra of the kinetic Taylor-Green vortex.  They may be caused by interferences between structures located near the symmetry planes that are separated by  a distance of $\pi$.  These oscillations can be removed by averaging the spectra on shells of width $\Delta k=2$ \cite{brachet83}.

We also verified through visualizations that the physical structures that appear in the flow are nearly identical between the two runs.  Therefore, we expect no difference in behavior in the ideal case for a full run without imposed symmetries up to the resolution and time achieved here and conclude that a run with imposed symmetries gives the same results at a substantially lower cost.

\section{Conclusion and perspectives} \label{s:conclu}

The existence of singularities of fluid equations has been a subject of many studies employing symmetric flows, since evidence of their occurrence in finite time is all that is needed to disprove the regularity of the equations, however constrained the flow in question may be.  In this sense, use of the IMTG flow proposed here is a well-accepted procedure, and it allows for reaching resolutions never attained before in MHD using spectral accuracy.

Using an IMTG initial configuration and maintaining the symmetries at all times, we found that the ideal flow appears to remain regular throughout the interval of reliability, although an acceleration of the exponential decay occurs just before the limit of the resolution is reached in the run corresponding to a regular grid of $2048^3$ points.  Our conclusion is based both on examination of the temporal evolution of current and vorticity norms and on the logarithmic decrement fitting technique.  Higher resolution runs will have to be performed to check later time behavior---in particular to confirm whether a rotational discontinuity spontaneously appears.  The study does not preclude the occurrence of singularities at later times, nor does it preclude their occurrence in other types of ideal MHD flows, as considered in pioneering studies making use of finite difference adaptive mesh refinement \cite{grauer2} or specific configurations of a weakly compressible flow \cite{kerr_brandenburg}.  Another possibility for computing longer in a DNS might be to use a filter for small scales different from the standard 2/3 dealiasing rule, as proposed in \cite{hou}.

Apart from the singularity issue, we saw an acceleration of the dynamics never observed before in this type of study, either in an Euler flow or in ideal MHD.  This phenomenon can be linked specifically to the geometric configuration of two current sheets becoming thinner individually in the presence of a large-scale shear and, at the same time, accelerating toward each other because of a differential in magnetic pressure.

The next obvious step is to compute the turbulent decay of the IMTG flow at a magnetic Prandtl number of unity and examine the interactions between turbulent eddies and Alfv\'en waves \cite{1536ab}; of particular interest is the potential emergence of non-Kolmogorovian energy spectra.  For example, it was predicted on a phenomenological basis by Iroshnikov and Kraichnan that Alfv\'en waves would slow the dynamics of MHD flows, relative to neutral fluids, and thereby produce a shallower spectrum; this shallower spectrum can also be derived using the Lagrangian renormalized approximation generalized to MHD \cite{yoshida07}.  Other spectral laws have been derived, for instance, in the context of anisotropic (``weak'') MHD turbulence; scaling laws depending on the characteristic timescales of the system may emerge as well, such as when forcing functions are introduced with their intrinsic correlation times in the presence of a uniform strong magnetic field \cite{MHDspec}.  By using the IMTG flow and implementing numerically its symmetries to be able to attain substantially higher numerical resolutions, we can continue the search for scaling laws in MHD.

Finally, the properties of the second TG flow (corresponding to the conducting case) will also be examined in a future work.  Since rotational structures, including double discontinuities \cite{whang}, have been observed in interplanetary space in a much more complex environment involving compressibility and Hall currents, the flows proposed here may prove useful in understanding the interplay between turbulent eddies, waves, and small-scale structures in interplanetary space and the heliosphere.

\vskip0.2truein

{\it Computer time was provided through NSF MRI -CNS-0421498, 0420873 and 0420985, as well as NCAR, the University of Colorado, and an IBM Shared University Research (SUR) program grant. E.L. was supported in part from an NSF IGERT Fellowship (Joint Program in Applied Mathematics and Earth and Environmental Science) at Columbia. P.D.M. is a member of the Carrera del Investigador Cient\'{\i}fico of CONICET. Finally, visualizations use the VAPOR software for analysis of terascale datasets \cite{vapor}.}

\end{document}